\documentclass[conference]{IEEEtran}
\IEEEoverridecommandlockouts
\usepackage[ruled,vlined]{algorithm2e}
\usepackage{enumitem}
\usepackage{array}
\usepackage{wrapfig}
\usepackage{amsmath,amsfonts}
\usepackage[noend]{algpseudocode}
\usepackage{graphicx}
\usepackage{textcomp}
\usepackage{float}
\usepackage{listings}
\usepackage{xspace}
\usepackage{multirow}
\usepackage{amsthm}

\usepackage{balance}
\usepackage{algpseudocode}
\usepackage{colortbl}

\usepackage[skins]{tcolorbox}
\usepackage{xcolor,pifont}
\usepackage{multicol}
\newcommand*\colourcheck[1]{%
	\expandafter\newcommand\csname #1check\endcsname{\textcolor{#1}{\ding{52}}}%
}
\colourcheck{blue}
\colourcheck{green}
\colourcheck{red}
\newtcolorbox{boxB}[2][]{%
  enhanced,colback=white,colframe=black,coltitle=black,
  sharp corners,
  toprule=1.0pt,
  rightrule=0.3pt,
  leftrule=0pt,
  bottomrule=0pt,
  fonttitle=\itshape\scshape\large,
  left=0pt,right=5pt,top=5pt,bottom=3pt,
  attach boxed title to top right={yshift=-0.3\baselineskip-0.4pt,xshift=-5mm},
  boxed title style={tile,size=minimal,left=0.2mm,right=0.5mm,
    colback=white,before upper=\strut},
  title=#2,#1
}

\newcommand{\tool}{\textsc{DualMine}\xspace}

\def\BibTeX{{\rm B\kern-.05em{\sc i\kern-.025em b}\kern-.08em
    T\kern-.1667em\lower.7ex\hbox{E}\kern-.125emX}}

\newboolean{showcomments}
\setboolean{showcomments}{true}
\ifthenelse{\boolean{showcomments}}
 { \newcommand{\mynote}[2]{
      \fbox{\bfseries\sffamily\scriptsize#1}
        {\small$\blacktriangleright$\textsf{\emph{#2}}$\blacktriangleleft$}}}
        { \newcommand{\mynote}[2]{}}

\usepackage{tikz}

\newcolumntype{L}[1]{>{\raggedright\arraybackslash}p{#1}}

\newcommand{\code}[1]{{\footnotesize\texttt{#1}}}
\usepackage{amsthm}
\definecolor{dkgreen}{rgb}{0,0.6,0}
\definecolor{gray}{rgb}{0.5,0.5,0.5}
\definecolor{lightgray}{rgb}{211, 211, 211}
\definecolor{mauve}{rgb}{0.58,0,0.82}

\definecolor{custom-red}{rgb}{0,0,1}
\definecolor{custom-blue}{rgb}{0,0,1}

\definecolor{c1}{HTML}{f4cccc}
\definecolor{c2}{HTML}{f5cdcd}
\definecolor{c3}{HTML}{fffcfc}
\definecolor{c4}{HTML}{ffffff}
\definecolor{c5}{HTML}{ffffff}
\definecolor{c6}{HTML}{fffdfd}
\definecolor{c7}{HTML}{f5cfcf}
\definecolor{c8}{HTML}{fffbfb}
\definecolor{c9}{HTML}{ffffff}
\definecolor{c10}{HTML}{fffdfd}
\definecolor{c11}{HTML}{fefafa}
\definecolor{c12}{HTML}{fef7f7}
\definecolor{c13}{HTML}{ffffff}
\definecolor{c14}{HTML}{fffefe}
\definecolor{c15}{HTML}{ffffff}
\definecolor{c16}{HTML}{fefafa}
\definecolor{c17}{HTML}{fdf3f3}
\definecolor{c18}{HTML}{fffefe}
\definecolor{c19}{HTML}{fdf5f5}
\definecolor{c20}{HTML}{ffffff}

\makeatletter
\newcommand{\linebreakand}{%
  \end{@IEEEauthorhalign}
  \hfill\mbox{}\par
  \mbox{}\hfill\begin{@IEEEauthorhalign}
}
\makeatother

\begin{document}

\title{{\tool}: Static-Dynamic REST API Constraint Discovery with Dual Validation}

\author{
\IEEEauthorblockN{Tu Nguyen}
\IEEEauthorblockA{\textit{University of Science, VNU-HCM}\\
Ho Chi Minh City, Vietnam\\
23C15017@student.hcmus.edu.vn}
\and
\IEEEauthorblockN{Huy Nguyen}
\IEEEauthorblockA{\textit{University of Science, VNU-HCM}\\
Ho Chi Minh City, Vietnam\\
22127154@student.hcmus.edu.vn}
\and
\IEEEauthorblockN{Juan Carlos Alonso Valenzuela}
\IEEEauthorblockA{\textit{Universidad de Sevilla}\\
Seville, Spain\\
javalenzuela@us.es}
\and
\IEEEauthorblockN{Thanh Nguyen}
\IEEEauthorblockA{\textit{University of Science, VNU-HCM}\\
Ho Chi Minh City, Vietnam\\
22127389@student.hcmus.edu.vn}
\and
\IEEEauthorblockN{Tien N. Nguyen}
\IEEEauthorblockA{\textit{University of Texas at Dallas}\\
Richardson, TX, USA\\
tien.n.nguyen@utdallas.edu}
\and
\IEEEauthorblockN{Vu Nguyen\textsuperscript{*}}
\IEEEauthorblockA{\textit{University of Science, VNU-HCM; Katalon LLC.}\\
Ho Chi Minh City, Vietnam\\
nvu@fit.hcmus.edu.vn}
\thanks{\textsuperscript{*}Corresponding author.}
}

\maketitle
\thispagestyle{plain}

\begin{abstract}
REST API constraints capture semantic properties of API responses and are essential for automated test oracle generation, but they are difficult to discover reliably. Static approaches infer constraints from API specifications and documentation, but their results may be affected by incomplete, ambiguous, or outdated specifications. Dynamic approaches mine invariants from execution traces, but their results depend on execution coverage and may include coincidental properties that hold only for the observed executions. 
This paper presents {\tool}, a hybrid framework for REST API constraint discovery that integrates specification-based constraint mining with runtime invariant mining. It first extracts candidate constraints from OpenAPI specifications using an LLM-based static miner and from request-response traces using dynamic invariant mining. It then performs asymmetric dual validation: runtime evidence is used to validate or refute specification-derived constraints, while specification-aware LLM reasoning is used to filter implausible log-derived invariants~without discarding plausible undocumented behaviors. Finally, it applies counterexample-guided refinement by performing targeted API executions to resolve uncertain, overlapping, or conflicting constraints. We evaluate {\tool} on 39 real-world REST APIs and compare it against state-of-the-art static-only, dynamic-only, and constraint discovery approaches. The results show that it improves the quality of discovered constraints by reducing unsupported constraints, retaining complementary constraints missed by individual approaches, which helps detect 48 real REST API faults.

%for REST API test oracle generation.

% constraints.

%REST API constraints are essential for automated testing and API validation. Existing discovery techniques rely on either static information from API specifications or dynamic information from execution traces. While static approaches provide semantic insights but may be affected by incomplete documentation, dynamic approaches capture actual behavior but are limited by execution coverage and coincidental invariants.

%This paper presents {\tool}, a hybrid REST API constraint discovery framework that combines the complementary strengths of static and dynamic analysis. {\tool} first extracts candidate constraints from API specifications using an LLM-based static analysis component and from execution traces using dynamic invariant mining. It then performs dual validation, where runtime evidence is used to verify specification-derived constraints, while LLM-based semantic reasoning filters implausible dynamically mined invariants. To further improve discovery quality, {\tool} incorporates a counterexample-guided refinement loop that actively generates new API executions to challenge uncertain constraints and iteratively refine the discovered constraint set.

%{\tool} consistently achieves higher precision and coverage than existing techniques by reducing false positives while uncovering constraints missed by individual approaches.
\end{abstract}

%\begin{IEEEkeywords}
%Keywords
%\end{IEEEkeywords}
\section{Introduction}

% Refer to AGORA~\cite{alonso2023agora,Alonso2025AGORA_plus}.

%oracle generation, fault detection, and specification validation. These constraints

%REST API {\em test oracles (i.e., constraints)} play a crucial role in automated~testing because they capture semantic relations among API inputs and outputs that cannot be expressed solely through response schemas or status codes. Prior REST API oracle work has explored metamorphic relations as an alternative oracle mechanism when explicit expected outputs are unavailable~\cite{segura2018metamorphic}. Thus, accurately discovering API constraints is crucial for verifying the logical correctness of API behaviors.}
REST API test oracles (i.e., constraints) play a crucial role in automated testing as they capture semantic relations among API inputs and outputs that cannot be expressed solely through
response schemas or status codes. Thus, accurately discovering API constraints is crucial for verifying the logical correctness of API behaviors.
Existing approaches to REST API constraint discovery typically rely on a single source of information. Static approaches (e.g., SATORI \cite{alonso2025satori} and RBCTest~\cite{huynh2026rbctest}) infer constraints from API specifications and schemas. While~they can exploit rich information documented by API developers, their effectiveness is limited by incomplete, ambiguous, or outdated specifications. As a result, many inferred constraints may not reflect the actual behavior of the implemented service. In contrast, dynamic approaches (e.g., AGORA/AGORA+~\cite{alonso2023agora,Alonso2025AGORA_plus}) discover constraints from execution observations. Although they capture real system behaviors, they are inherently dependent on execution coverage and often generate coincidental invariants that hold only for the observed executions.

The limitations of static and dynamic constraint discovery approaches are largely complementary, suggesting a strong opportunity for integration. Static approaches can leverage API specifications, descriptions, and schemas to infer semantically meaningful constraints, including those associated with rarely exercised behaviors. Consequently, they can compensate for the limited execution coverage of dynamic techniques. However, because specifications may be incomplete, ambiguous, or outdated, static methods may infer constraints that do not accurately reflect the implemented service. Dynamic approaches address this weakness by deriving constraints from actual API executions, thereby providing evidence of real system behavior and helping validate or refute statically inferred constraints. At the same time, dynamic methods are susceptible to generating coincidental invariants due to insufficient coverage or biased observations. The semantic knowledge extracted by static analysis can help identify and eliminate such spurious constraints while also guiding exploration toward uncovered behaviors. Therefore, the strengths of each approach naturally mitigate the weaknesses of the other: {\em static analysis provides semantic completeness and generalization, whereas dynamic analysis provides behavioral validation and implementation-level accuracy}. Integrating these complementary sources of evidence offers the potential to discover API constraints that are both semantically meaningful and empirically valid. 
%
%{\color{custom-blue} 
%This complementarity follows a broader software-testing lineage in which concrete execution, symbolic reasoning, and invariant inference are combined to improve exploration and reduce spurious conclusions~\cite{godefroid2005dart,majumdar2007hybrid,csallner2008dysy}.
%}

These complementary strengths and weaknesses motivate {\tool}, a {\bf \em hybrid approach} that integrates static and dynamic evidence. 
{\tool} first extracts candidate constraints from API specifications and execution logs through static and dynamic mining, respectively. 
However, {\em simply unioning specification-derived constraints and
runtime-derived invariants is {\bf insufficient}}: the two sources may produce
duplicates, partially overlapping constraints, or conflicting hypotheses,
and each source may also produce constraints that cannot be confirmed
from the other.
Therefore, {\tool} applies a {\bf \em dual-validation process} in which execution evidence is used to validate specification-derived constraints, while LLM reasoning is used to identify and filter semantically implausible runtime invariants. 
%
%{\color{custom-blue}
%Because LLM-generated hypotheses may be unsupported or hallucinated~\cite{ji2023hallucination}, 
To improve reliability, it introduces a counterexample-guided refinement ~\cite{clarke2003cegar}, to generate new API executions that challenge uncertain constraints and refine the discovered constraint set.
By integrating complementary sources of knowledge and validation evidence, {\tool} reduces false positives while improving the completeness of discovered constraints over the baselines. 
%

%demonstrate that the proposed framework consistently outperforms static-only and dynamic-only approaches in constraint mining.

In brief, this paper makes the following contributions:

1. {\tool}: a hybrid framework integrating LLM-based static and dynamic constraint mining for API testing.

2. A dual-validation and counterexample-guided refinement mechanism that
combines execution evidence and LLM-based semantic reasoning to improve
constraint reliability.

3. An empirical evaluation to compare {\tool} against the state-of-the-art static and dynamic baselines. Importantly, {\tool} {\bf \em revealed 48 bugs in real-world REST APIs}.
%the constraint discovery capability of {\tool} with that of the individual approach and with the state-of-the-art approaches.

4. A publicly available replication package and {\tool} tool including the source code and the data~\cite{dual_mine_replication_2026}.

\section{Motivation}
\label{sec:motiv}

%\subsection{Motivating Example}
%\label{sec:motiv_example}

We use GitLab’s API to illustrate the challenges of constraint mining in REST API testing. Consider the \code{ssh\_url\_to\_repo} field returned by the GitLab project creation endpoint. RBCTest~\cite{huynh2026rbctest}, a LLM-based static mining approach extracts specification-driven constraints such as \code{isRegex(return.ssh\_url\_to\_repo}, \code{`$^{}git@.*$')} and \code{contains(return.ssh\_url\_to\_repo, `.git')}, capturing formatting rules like the SSH prefix and required suffix.

In contrast, dynamic mining can uncover deep semantic relationships across fields. For example, AGORA+~\cite{Alonso2025AGORA_plus} reveals that \code{return.path\_with\_namespace} is a substring of \code{return.ssh\_url\_to\_repo}, reflecting how the repository path is embedded within the SSH URL. This dependency is not stated in the specification and can only be observed through actual API executions.
In brief, while static mining captures format- and pattern-based constraints, dynamic mining is able to reveal richer inter-field dependencies that are valuable for generating valid test inputs and detecting API inconsistencies.

%We use GitLab's Branch API to illustrate the cha llenges of constraint mining for REST API testing. Consider the \code{ssh\_url\_to\_repo} property returned by the GitLab project creation endpoint. Static mining extracts specification-derived constraints such as \code{isRegex(return.ssh\_url\_to\_repo, '^{}git@.*$')} and \code{contains(return.ssh\_url\_to\_repo, '.git')}, indicating that the SSH repository URL must start with \code{git@} and contain the \code{.git} suffix. In contrast, dynamic mining discovers the cross-field constraint that \code{return.path\_with\_namespace} is a substring of \code{return.ssh\_url\_to\_repo}. This semantic relationship cannot be directly inferred from the API specification and only emerges from observing actual API executions. Therefore, while static mining captures format- and pattern-based constraints, dynamic mining is able to reveal richer inter-field dependencies that are valuable for generating valid test inputs and detecting API inconsistencies.

This example highlights the complementary strengths and weaknesses of existing approaches. Static techniques such as SATORI~\cite{alonso2025satori} and RBCTest can effectively extract constraints explicitly documented in API specifications, including type, format, and schema-related restrictions. However, they are less reliable at inferring semantic relationships among fields that are not specified in the API documentation. Conversely, dynamic approaches such as AGORA+~\cite{Alonso2025AGORA_plus} can discover such runtime relationships from execution traces, but their effectiveness depends on the quality and coverage of the observed executions.
Consequently, static approaches may miss hidden semantic constraints, while dynamic approaches may fail to infer them when insufficient runtime evidence is available.

However, neither source is reliable by itself. A specification-derived
format constraint may be outdated if the implementation returns a
different URL variant, while a dynamically mined constraint may be
coincidental if all observed repositories are from the same namespace. Thus, a framework~must not only combine constraints
from the two sources, but also {\em validate them against complementary
evidence and generate extra executions when the available evidence
is insufficient}.

\section{Key Ideas}
\label{sec:motiv_ideas}

We design our solution with the following key ideas.

\textbf{\textit{Key Idea 1: A hybrid static-dynamic framework for REST API constraint discovery.}}
The above observations motivate a~hybrid constraint mining solution that combines specification-based and runtime-based approaches to achieve the best of both worlds for more~comprehensive constraint extraction.
We propose {\tool}, a hybrid framework that combines specification-derived constraints and runtime invariants to improve the discovery of REST API constraints. The static component leverages API specifications and LLM reasoning to infer candidate constraints, while the dynamic component mines invariants from execution logs to capture runtime behaviors that are not explicitly documented. By integrating two complementary sources, we expect to achieve higher discovery accuracy than relying on either static or dynamic analysis individually.

\textbf{\textit{Key Idea 2: A dual-validation framework combining execution evidence and LLM reasoning.}}
We introduce a dual-validation framework that leverages complementary validation strategies for constraints derived from two different sources. For the specification-derived constraints, execution logs are used as runtime evidence to verify their consistency with observed API behaviors. For the dynamically mined invariants, LLM reasoning is employed to identify and filter spurious constraints that arise from coincidental patterns in execution traces. 
For example, in the GitLab Branch API, a static approach may infer from the specification that the returned
\code{ssh\_url\_to\_repo} field must start with the SSH prefix \code{git@} and end with the suffix \code{.git}. {\tool} validates such constraints against actual API responses to confirm that they hold in the implemented service. Conversely, a dynamic approach may infer that a field always takes a particular constant value because all observed executions happen to come from the same project or user. In this case, LLM-based semantic reasoning can recognize that the invariant is likely coincidental rather than a true API constraint and filter it out.
With the dual-validation via the combination of execution-based and semantic-based reasoning, the framework improves the reliability and precision of discovered constraints.

% \begin{figure}[t]
% \centering
% \includegraphics[draft,width=\textwidth]{figures/architecture.pdf}
% \caption{Architecture}
% \end{figure}

\begin{figure*}[t]
  \centering
  \includegraphics[width=0.9\textwidth]{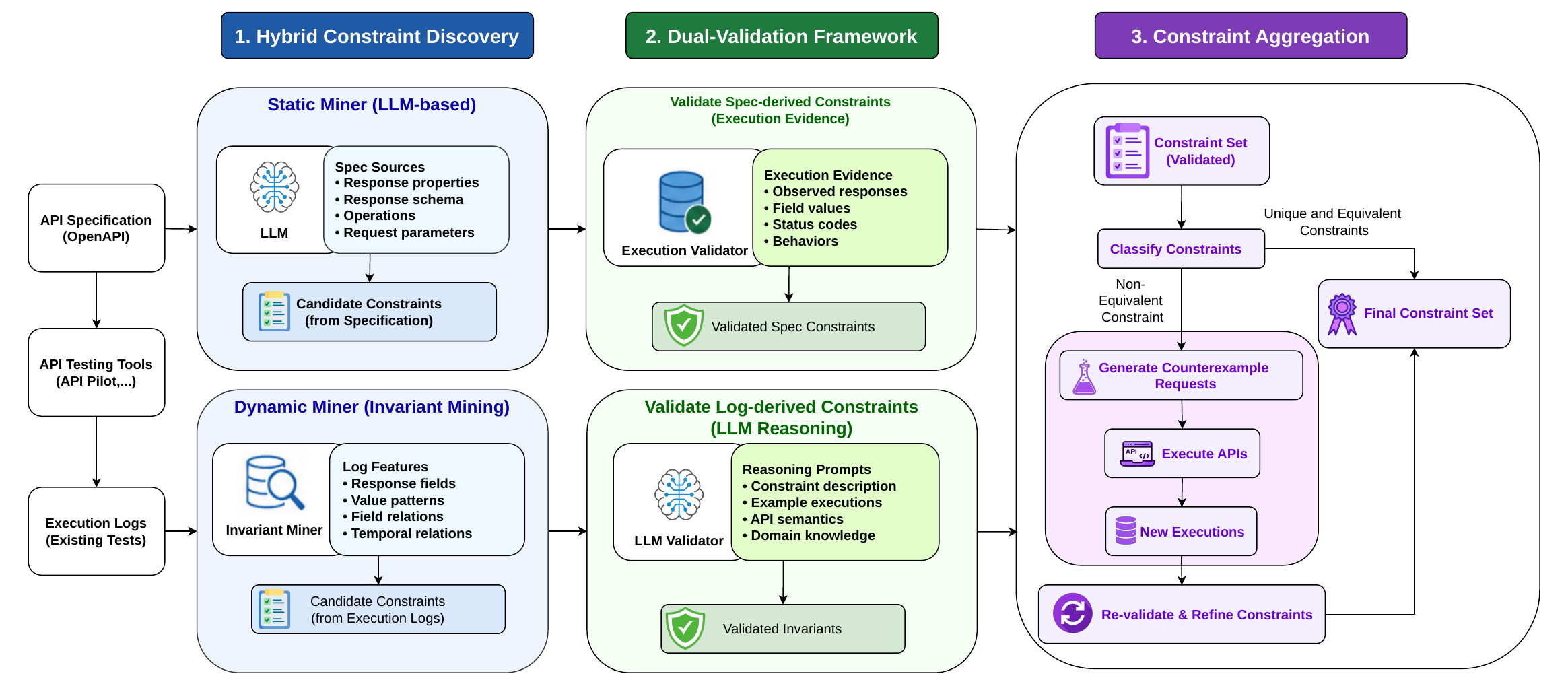}
  \vspace{-6pt}
  \caption{{\tool}: Static-Dynamic REST API Constraint Discovery with Dual Validation}
  \label{fig:architecture}
\end{figure*}

\textbf{\textit{Key Idea 3: A counterexample-guided refinement for REST API constraints.}}
We propose a counterexample-guided refinement that iteratively improves the quality of discovered constraints.
%
%{\color{custom-blue}
%The loop borrows the abstract check-refine pattern from counterexample-guided abstraction refinement~\cite{clarke2003cegar}, but {\tool} does not perform formal model checking; instead, it generates targeted REST API executions to resolve uncertain or conflicting constraint hypotheses.
%
When neither existing logs nor specification semantics can resolve a
constraint, {\tool} actively generates targeted executions to distinguish
competing hypotheses. Specifically, when a candidate constraint cannot be confidently validated, the framework generates API requests aimed at producing counterexamples to expose unexplored behaviors and gather additional execution evidence. The newly observed executions are then used to validate, refine, or discard the candidate constraints. 
For example, suppose dynamic mining infers that
\code{return.path\_with\_namespace} must always be a substring of
\code{return.ssh\_url\_to\_repo}. If this relationship is observed only
for a small number of projects, {\tool} actively generates
additional project-creation or branch-query requests with
different project names, namespaces, and visibility settings.
If the relationship continues to hold across these new executions,
the framework gains stronger evidence for accepting the
constraint. If a generated execution violates the relation,
the candidate constraint can be weakened, specialized to a
particular condition, or discarded. Through this iterative feedback process, we could reduce false positives while improving precision of constraint discovery.

%We propose a counterexample-guided refinement loop that continuously improves the quality of discovered constraints. When a candidate constraint cannot be sufficiently validated, the framework generates counterexample-oriented API requests to explore unexplored behaviors and collect additional execution evidence. The newly observed executions are then used to validate, refine, or discard candidate constraints. This iterative process reduces false positives and improves both the precision and coverage of REST API constraint discovery.

\section{{\tool} Architecture}
\label{sec:architecture}

\subsection{Overview}

Figure \ref{fig:architecture} illustrates the overall architecture of the proposed framework. The framework aims to automatically discover, validate, and refine behavioral API constraints by integrating information from statically mined API specifications and actual runtime executions. The architecture consists of three major phases: \textbf{1) Hybrid Constraint Discovery, 2) Dual-Validation Framework, and 3) Constraint Aggregation}.

%---- Remove this algorithm
%\input{sections/architecture_alg}
%======================

%Algorithm~\ref{alg:framework} summarizes {\tool} workflow.
In {\bf \em the hybrid constraint discovery phase}, it generates candidate constraints from two complementary sources:
specifications and API executions. {\bf \em Our novelty lies in the validation
and aggregation process}: instead of simply combining the~results, the candidate constraints are normalized, cross-validated, compared, and refined using additional evidence. The design combines the strengths of specification-driven reasoning and execution-log analysis while mitigating their individual limitations via validation and iterative refinement.

%The technical novelty of DUALMINE lies in this reconciliation process: it does not treat static and dynamic miners as independent sources whose outputs are simply combined, but as complementary hypothesis generators whose outputs must be normalized, cross-validated, compared, and refined using additional evidence.

%Algorithm~\ref{alg:framework} summarizes our hybrid framework.
%In {\bf \em the discovery phase}, the framework generates candidate constraints from two complementary sources. \textsc{LLMMiner} infers specification-derived constraints from documentation, while \textsc{DynaMiner} extracts log-derived invariants from historical API executions. The resulting constraint sets are collected.

%are combined to maximize both semantic coverage and runtime evidence.

To do so, the framework applies a {\bf \em dual-validation strategy} to improve constraint quality. %Specification-derived constraints are validated against execution logs using \textsc{ExecValidator} to remove unsupported hypotheses introduced by LLM reasoning. 
Specification-derived constraints are checked against execution logs using \textsc{ExecValidator}, which classifies them as supported, contradicted, or unresolved based on available runtime evidence.
Thus, log-derived invariants are validated through \textsc{LLMValidator}, which rejects semantically implausible
log-derived invariants while retaining plausible undocumented behaviors
for further refinement.

%which filters spurious correlations that are not semantically supported by the API specification.

Finally, in the {\bf \em constraint aggregation phase}, the validated constraints are classified into unique, equivalent, and conflicting hypothesis sets. Two constraints are {\bf \em equivalent} if they refer to the same API field
and impose logically identical or mutually subsuming predicates.
They {\bf \em conflict} if they refer to the same field or relation but cannot
hold simultaneously, or if one is violated by evidence
supporting the other.
{\bf \em Unique} constraints are retained directly, while equivalent constraints are consolidated by selecting a representative constraint. 
The most executable and
specification-aligned form from each equivalent group is selected.
For each non-equivalent hypothesis set, our framework generate targeted {\bf \em counterexamples}, executes them against the target API, and resolves the competing hypotheses based on the new observations. The resulting~set contains validated, non-redundant, and representative API constraints.

\subsection{Hybrid Constraint Discovery}
The first phase generates candidate constraints from both static and dynamic sources.
Internally, {\tool} represents each candidate constraint as
\(c=\langle o, X, Y, \phi, src, E\rangle\), where \(o\) is the API
operation, \(X\) is the set of request parameters, \(Y\) is the set of
response fields, \(\phi\) is the predicate, \(src\in\{\textit{static},
\textit{dynamic}\}\) denotes the source, and \(E\) stores supporting
evidence such as specification text or observed executions. The predicate \(\phi\) is represented in a normalized executable form such as a comparison, membership constraint, format
predicate, substring relation, or input-output dependency.

\subsubsection{\bf \em Static Miner}
%The static miner derives candidate behavioral constraints from API specifications. Given an OpenAPI document, it extracts endpoint descriptions, request parameters, response schemas, and operation metadata, and leverages an LLM to infer input–output relationships. The resulting specification-derived constraints capture the intended API behavior encoded in the specification. However, since API specifications may be incomplete or ambiguous, the inferred constraints require further validation. 
In our hybrid framework, we build upon RBCTest~\cite{huynh2026rbctest} as the static constraint miner to extract candidate response-body constraints from OpenAPI specifications. RBCTest is an LLM-based static approach that analyzes operation descriptions, request parameters, response schemas, property descriptions, data types, formats, and examples to infer logical constraints that can serve as API test oracles after validation and refinement. Specifically, it mines two main types of constraints: request-response constraints, which capture how input parameters restrict or determine response properties, and response-property constraints, which capture value ranges, formats, enumerations, and semantic rules within response bodies. 
%To reduce hallucinated or weakly supported constraints, RBCTest employs an Observation-Confirmation prompting scheme, where the LLM first observes potential constraints from specification text and then confirms whether the inferred constraint is sufficiently grounded and useful for test generation.
%
%{\color{custom-blue}
%This design differs from non-LLM static REST API analyses such as ReSpecTor and REST$\pi$, which recover REST specifications or path-sensitive API types from implementations rather than mining response-body oracle constraints from an OpenAPI document~\cite{huang2024respector,aldrich2025restpi}.}
In our approach, these RBCTest-mined constraints form the static candidate set. They provide semantically meaningful constraints that may not be observable from limited executions, especially for rarely exercised fields or behaviors. 
For example, from the schema of a GitLab repository
field, the static miner may infer that \code{return.ssh\_url\_to\_repo}
must start with \code{git@} and end with \texttt{.git}.
{\tool} {\em normalizes} each RBCTest-mined constraint into a common
predicate representation that records the target operation, involved
request parameters, response fields, predicate expression, and static
evidence from the OpenAPI specification.

%However, because specification-derived constraints may be incomplete, ambiguous, or inconsistent with the implemented service, our hybrid framework further validates and refines them using runtime evidence and counterexample-guided executions.

\subsubsection{\bf \em Dynamic Miner}
%The dynamic miner derives candidate constraints from historical API execution logs. It analyzes runtime observations, including response values and inter-field relationships, and applies invariant mining techniques to identify recurring behavioral properties that consistently hold across executions. The resulting log-derived constraints capture empirically observed API behavior. However, because the discovered invariants depend on available execution traces, they may be incomplete or reflect spurious correlations under limited coverage.
We use AGORA+~\cite{Alonso2025AGORA_plus} as the dynamic constraint miner to extract candidate constraints from observed API executions. AGORA+ is a black-box oracle-generation approach that learns likely invariants from API requests and their corresponding responses. Given an OpenAPI specification together with a set of request-response traces, AGORA+ uses its Beet front-end to translate API inputs and outputs into Daikon-compatible declaration and trace files, and then applies a customized version of Daikon to mine invariants over request parameters and response fields. These invariants capture runtime-observed properties such as numerical bounds, array-size relationships, string formats, specific values, and inter-field relations. 
From observed request-response traces, the dynamic miner may infer that
\code{return.path\_with\_namespace} is a substring of
\code{return.ssh\_url\_to\_repo}.
{\tool} normalizes each AGORA+-mined invariant into the same
constraint representation, while preserving its runtime support,
including the number of executions in which the invariant was observed
and the fields involved in the invariant. Constraints that cannot be mapped to an executable predicate are retained as unresolved semantic constraints and are considered only during
LLM validation, not during execution-based~falsification.
%However, because dynamically mined invariants are learned from finite executions, they may include coincidental or overfitted properties caused by limited coverage or biased traces. Therefore,
%{\tool} further validates these invariants using specification-aware semantic reasoning and counterexample-guided executions.

\subsection{Dual-Validation Framework}

The key design principle is asymmetric validation: runtime
evidence is used to falsify specification-derived hypotheses, while
specification semantics is used to reject dynamically mined coincidences.
In Phase 2, the candidate constraints mined from the specification ($C_s$) and the execution logs ($C_d$) are inherently susceptible to distinct types of inaccuracies.
The execution logs may come from existing API test suites, API testing
tools, or executions generated during {\tool}’s refinement loop. Each
log entry records the HTTP request, concrete parameter values, response
status, and response body.
LLMs may hallucinate constraints based on generalized knowledge~\cite{ji2023hallucination}, while dynamic mining tools often overfit to the limited data available in execution logs. To address these complementary weaknesses, our framework employs a dual-validation mechanism. This phase cross-checks the outputs of each discovery method against the foundational data of the other: specification-derived constraints are evaluated empirically against execution logs, and log-derived constraints are evaluated semantically against the  specification. The union of these refined sets forms the validated constraint pool~$C_v$.

\subsubsection{\bf \em Validation of Specification-Derived Constraints}

Specification-derived constraints ($C_s$) extracted via the LLM frequently contain hallucinations, over-approximations, or unsupported behavioral assumptions that diverge from the actual implementation of the API. To mitigate this issue, our execution validator acts as an empirical filter by evaluating each candidate constraint against API execution logs ($L$).
A spec-derived constraint is classified as {\em supported, contradicted}, or {\em unresolved}. A constraint is supported if all relevant
executions satisfy it, contradicted if at least one successful execution
violates it, and unresolved if the logs do not contain sufficient
coverage for the involved parameters or response fields.

Specifically, the validation engine parses the recorded HTTP requests and responses in $L$ to extract the concrete runtime values of the parameters involved in each proposed constraint. 
Only successful executions whose responses contain all fields referenced by the constraint are considered relevant. Executions with error status codes, missing optional fields, or incompatible response schemas are handled separately and do not automatically falsify the constraint.
The constraint is then evaluated as a boolean predicate over these runtime values. If a constraint consistently evaluates to true across all relevant, successful executions, it is considered empirically sound and is retained. Conversely, if runtime evidence provides a direct contradiction—such as a successful `200 OK` response that violates an LLM-proposed dependency—the constraint is discarded as spurious. This empirical validation step filters constraints contradicted by
observed executions and records the degree of runtime support for the
remaining constraints.

%This empirical validation step efficiently filters out LLM hallucinations and ensures the discovered knowledge strictly aligns with the actual API runtime behavior, producing the refined set $C_s'$.

% Specification-derived constraints may contain hallucinations or unsupported behavioral assumptions introduced by the LLM. To mitigate this issue, each candidate constraint is evaluated against historical API executions. Constraints that consistently hold across observed executions are retained, while those contradicted by runtime evidence are discarded or refined. This validation step filters spurious constraints and aligns the discovered knowledge with actual API behavior.
\subsubsection{\bf \em Validation of Log-Derived Constraints}

Conversely, invariants mined dynamically from execution logs ($C_d$) often capture spurious statistical correlations rather than genuine API rules. This overfitting occurs primarily due to limited execution coverage in $L$, where coincidental relationships in a small data sample are mistakenly generalized into rigid rules. For example, from a limited set of GitLab project responses, an invariant miner might infer an arithmetic constraint such as $\code{input.id\_after} \ge \code{size}(\code{return}[])$, simply because the numerical values of the \code{id\_after} parameter happened to be larger than the array size of the returned projects in the specific log traces. However, the \code{id\_after} parameter is fundamentally a pagination cursor and is not semantically related to bounding the number of returned items.

To systematically eliminate such false positives, our LLM-based validator leverages the LLM's natural language reasoning capabilities to evaluate each dynamic invariant against the OpenAPI specification ($S$). The LLM validator does not require every dynamic invariant to be
explicitly stated in the specification. Instead, it evaluates if
the invariant is semantically plausible given the endpoint description,
parameter meanings, response schema, and domain context. Invariants
that contradict the specification or lack any plausible semantic
connection are rejected, while plausible but undocumented invariants
are retained for further refinement.

For each invariant in $C_d$, we construct a verification prompt containing the inferred rule alongside its corresponding API documentation, including parameter descriptions, endpoint semantics, and response schemas. The LLM is tasked with~deciding whether the constraint represents a logical, semantically sound rule or merely an accidental artifact of the data. 
%Invariants lacking clear semantic justification in the specification are discarded. 
Invariants that contradict the specification or lack any plausible semantic connection to the endpoint behavior are discarded.
By anchoring dynamic invariants in the text semantics of~the documentation, this eliminates overfitting and much improves the precision of the refined log-derived constraints $C_d'$.

% To improve reproducibility, the LLM validator is prompted with a
% structured template and must return one of three labels:
% \code{Accept}, \code{Reject}, or \code{Uncertain}, together with
% the specification evidence or semantic rationale supporting the decision.
% For dynamic invariants, those labels correspond to plausible, implausible, and unresolved
% validation outcomes,~respectively.

% Invariants mined from execution logs may capture spurious correlations caused by limited execution coverage rather than genuine API behavior. For example, from a set of GitLab project responses, an invariant miner may infer a constraint such as \code{input.id\_after ≥ size(return[])}, even though the parameter id\_after is not semantically related to the number of returned projects. To mitigate such false positives, we employ an LLM-based validator that evaluates each invariant against API documentation, parameter descriptions, endpoint semantics, and response schemas. Invariants lacking semantic support are discarded, while semantically justified invariants are retained. This step reduces overfitting to observed executions and improves the precision of the discovered constraints.

\subsection{Constraint Aggregation and Refinement}

For each retained constraint in \(C'_s \cup C'_d\), {\tool} records a support score capturing the number of relevant executions, the number of satisfying
executions, and the validation outco\-me: if the constraint is
supported, contradicted, or unresolved.

After dual validation, {\tool} obtains the validated constraint pool
\(C_v = C'_s \cup C'_d\), which contains specification-derived
constraints and log-derived invariants that have survived their
respective validation steps. However, since historical logs ($L$) are passive
and specifications may be incomplete, \(C_v\) may still contain
redundant, overlapping, unresolved, or partially incorrect constraints.
Phase 3 addresses this limitation through constraint classification and
counterexample-guided refinement. This phase transitions the framework from passive observation to proactive exploration by categorizing candidate constraints and actively probing the target API to detect equivalent or non-equivalent~constraints.

%resolve remaining ambiguities.

%{\color{red}REWRITE.}
% {\color{red}REWRITE with MORE DETAILs. Also, need to refer to Algorithm 1.}

% Although dual validation significantly improves precision, some constraints may remain incorrect, overly general, or redundant. Therefore, the framework introduces a counterexample-guided refinement phase.
% The process begins with the set of validated constraints.

\subsubsection{\bf \em Constraint Classification}

The refinement process begins by passing the validated constraint pool ($C_v$) into the constraint classification procedure. This module organizes the candidate constraints into three distinct categories based on their semantic relationships and sources:

%\begin{itemize}

$\circ$ \textbf{Unique Constraints ($C_u$):} 
Unique constraints are the ones without an equivalent or
non-equivalent counterpart in the validated pool. They may originate from
either source, but are retained only when their validation evidence is~sufficient.
In our implementation, evidence is sufficient when the constraint is not
contradicted and either has runtime support above a threshold or is
semantically accepted by the LLM validator.

%Constraints originating from a single source (either solely from the specification $C_s'$ or solely from the execution logs $C_d'$) that do not overlap or conflict with any other rules.

$\circ$ \textbf{Equivalent Constraints ($C_e$):} Groups of constraints that assert the exact same logical rule for a given API property, despite being mined from different sources or expressed via disparate syntaxes. For example, the specification miner might extract a constraint in text (e.g., {\em ``age must be non-negative''}), while the dynamic miner infers a concrete arithmetic invariant (e.g., \code{input.age $\ge$ 0}). Retaining both definitions creates semantic redundancy and introduces unnecessary computational overhead in testing tasks.
{\tool} treats two constraints as equivalent when they refer to the
same operation and target field and impose the same predicate after
normalizing syntax, comparison direction, constants, and field aliases.

%In our implementation, \textsc{SelectRepresentative} prioritizes the constraint form that is most executable, precise, and traceable to evidence. When two equivalent forms are equally precise, DUALMINE prefers the specification-aligned form to preserve compatibility with the OpenAPI schema.

To resolve this issue, {\tool} invokes
\textsc{SelectRepresentative}$(C_e)$ to select one representative
constraint from each equivalent group. In our implementation,
\textsc{SelectRepresentative} prioritizes the constraint form that is
most executable, precise, and traceable to evidence. When two equivalent
forms are equally precise, we prefer the specification-aligned
form to preserve compatibility with the OpenAPI schema. These chosen
representatives, along with the unique constraints \(C_u\), are then
seeded directly into the final set \(C^*\).

%To resolve this issue, the framework invokes $\textsc{SelectRepresentative}(C_e)$ to act as an arbiter, electing either the \textbf{specification-derived} or the \textbf{runtime-derived} constraint to serve as the definitive representative for that target property. In our implementation, this function prioritizes the specification-derived format to ensure seamless compliance and backward-integration with the original OpenAPI schema. These chosen representatives, along with the unique constraints ($C_u$), are then seeded directly into the final constraint set ($C^*$).

$\circ$ \textbf{Non-equivalent Hypotheses ($H$):} Sets of constraints that exhibit source discrepancies or mutually exclusive behavioral assertions. For automated resolution, these hypotheses are sub-classified into four basic set-theoretic relationships based on their parameter domains and condition boundaries: \textit{subset} (one rule is a stricter variation of another), \textit{intersection} (rules overlap only under specific conditions), \textit{independence} (rules assert entirely incompatible behaviors for the same parameters), and \textit{union} (rules represent competing generalizations of the same observed behavior). These relationships are determined after grouping constraints by
operation and target fields and comparing their normalized predicates.
When logical~comparison is not directly possible, {\tool} asks the LLM
to classify the relationship using a constrained output schema.
E.g., a spec-derived constraint may say that
\code{return.status} belongs to \{\code{open}, \code{closed},
\code{merged}\}, while a dynamic one observes only
\{\code{open}, \code{closed}\}. This forms a subset relation: the
dynamic invariant is stricter but may reflect limited execution
coverage.

%\end{itemize}

% Validated constraints are classified into three categories: \textit{unique} constraints originating from a single source (specification or runtime); \textit{equivalent} constraints, which are logically identical and merged to reduce redundancy; and \textit{potentially conflicting} constraints. The latter exhibit source discrepancies and are further sub-classified into four set-theoretic relationships—subset, intersection, disjoint, and union—for further investigation.
\subsubsection {\bf \em Counterexample Generation}

As in the iterative loop, each non-equivalent hypothesis set $h \in H$ undergoes rigorous automated testing. For an uncertain or non-equivalent rule, we generate counter examples on $(h, S)$. Guided by the structural boundaries defined in the OpenAPI specification ($S$), the LLM acts as a targeted test generator.

Rather than generating random inputs, {\tool} generates
hypothesis-discriminating tests: requests whose expected outcomes
would satisfy one candidate constraint while violating another.
Generated requests are required to satisfy the OpenAPI schema and known
operation dependencies unless the hypothesis explicitly concerns invalid
inputs or error responses.
%Instead of generating random inputs, the LLM employs a differential testing paradigm. 
It analyzes the semantic differences within the hypothesis set $h$ and generates a suite of counterexample test cases ($T$) specifically designed to challenge the contested boundaries of the constraints. 
%The primary objective is to construct inputs that lie within the symmetric difference of the competing hypotheses, 
The goal is to generate requests likely to exercise the behavioral
region where the competing hypotheses make different predictions,
thereby maximizing the likelihood of violating at least one of the invalid rules and exposing its empirical boundary.

% For each uncertain or conflicting constraint, the framework automatically generates counterexample requests designed to challenge the constraint. The generation process aims to identify inputs that maximize the likelihood of violating the current hypothesis.

\subsubsection {\bf \em Targeted API Execution}

Once the targeted test suite \(T\) is generated, the framework passes
these inputs to the execution engine via
\(L_{new} \leftarrow \textsc{Execute}(A,T)\), where \(A\) denotes the
live target API.
When resolving constraints over successful response bodies, {\tool}
prioritizes requests expected to produce valid 2xx responses.
Requests that intentionally trigger error responses are used only when
the candidate constraint concerns error behavior or status-dependent
schemas.

Unlike historical logs that reflect nominal user behavior, this targeted
execution forces the API into boundary states and rarely exercised
behaviors. The resulting runtime observations are captured in a fresh
set of execution logs \(L_{new}\), which records how the API handles the
synthesized cases.

%Once the targeted test suite ($T$) is generated, the framework passes these inputs to the execution engine via $L_{new} \leftarrow \textsc{Execute}(A, T)$, where $A$ denotes the live target API. Unlike historical logs that reflect nominal user behavior, this targeted execution forces the API into boundary states and error-handling paths that are rarely traversed during normal operation. The resulting runtime observations are captured in a fresh, highly diverse set of execution logs ($L_{new}$), which explicitly records how the API handles the synthesized edge cases.

%When resolving constraints over successful response bodies, DUALMINE prioritizes requests expected to produce valid \(2xx\) responses. Requests that intentionally trigger error responses are used only when the candidate constraint concerns error behavior or status-dependent schemas.

% Generated requests are executed against the target API, producing new runtime observations. These executions expand behavioral coverage beyond existing test suites and expose previously unseen API behaviors.

\subsubsection {\bf \em Constraint Re-validation}
%{\color{custom-blue}{
%In the final step of the loop, the framework executes $\textsc{LLMResolveHypothesis}(h, L_{new})$ to analyze the newly acquired runtime evidence. By cross-referencing the HTTP response statuses and payload structures in $L_{new}$ against the conflicting assertions in $h$, the system systematically arbitrates the deadlock.

In the final step, the framework executes
\code{LLMResolveHypothesis}($h, L_{new}$) to analyze the newly
acquired runtime evidence. The resolver returns a set \(C_h\) of
resolved constraints, which may contain an accepted original constraint,
a weakened constraint, a specialized conditional constraint, or no
constraint if the evidence is insufficient. The final constraint set is
then updated as \(C^* \leftarrow C^* \cup C_h\).

By cross-referencing the HTTP response statuses and payload structures
in \(L_{new}\) against the non-equivalent assertions in \(h\), the system
resolves the conflict among competing hypotheses.

Under this empirical scrutiny, invalid hypotheses are falsified and discarded, while valid rules are confirmed. If a hypothesis requires adaptation to accommodate the new runtime evidence, it is weakened or specialized. 
%The resolved, verified constraint ($c$) is then extracted; if it is non-empty ($c \neq \emptyset$), it is appended to the final constraint set ($C^*$). 
The returned set \(C_h\) is added to \(C^*\); if the evidence is
insufficient, \(C_h\) is empty and the hypothesis remains unresolved or
is discarded according to the refinement budget.
This systematic loop repeats for all non-equivalent hypotheses in $H$. The culmination of this process is the return of the Final Constraint Set ($C^*$), which comprises a non-redundant constraint set that is better supported by both specification semantics and runtime evidence.

%an optimized, non-redundant, and highly precise representation of the true API behavioral constraints.

% As new observations are integrated into the validation framework, existing constraints are systematically confirmed, weakened, specialized, or discarded. This iterative refinement continues until convergence is achieved—specifically, when additional counterexamples no longer significantly alter the constraint set. The culmination of this process is the Final Constraint Set, which comprises validated, non-redundant, and behaviorally representative API constraints.
\section{Empirical Evaluation}
\label{sec:eval}
We conduct experiments to answer the following questions:

\textbf{RQ1 (Effectiveness).} How effective is \tool\ in discovering valid API behavioral constraints?

\textbf{RQ2 (Dual-Validation Capability).} How effectively does the dual-validation framework mitigate LLM hallucinations and filter out log-derived false positives?

\textbf{RQ3 (Counterexample-Guided Refinement).} To what extent does the counterexample-guided refinement improve constraints?

\textbf{RQ4 (Cost-Effectiveness Analysis).} What are the token costs associated with \tool?

%\textbf{RQ5 (Fault Detection).} How do {\tool}'s result help in fault detection?

\textbf{RQ5 (Fault Detection).} Can {\tool}'s constraints serve as test oracles for detecting real REST API faults?

\subsubsection*{Baselines}
We compared {\tool} against three state-of-the-art methods. For static constraint mining, we chose RBCTest~\cite{huynh2026rbctest} and SATORI~\cite{alonso2025satori}, while AGORA+~\cite{Alonso2025AGORA_plus} was selected as the representative baseline for dynamic mining. 
%The LLM used in our experiments is GPT-5, configured with minimal reasoning effort, to evaluate its raw performance. 
In our experiments, we used GPT-5, configured with minimal reasoning effort, as the underlying model.
We run all the tools five times and report mean value. 
% We also included GPT-5 (prompted with minimal reasoning effort) to evaluate LLM's raw performance.

\subsubsection*{Datasets}
We constructed a dataset comprising 39 REST API endpoints collected from seven real-world services previously used in RBCTest \cite{huynh2026rbctest} and AGORA+\cite{Alonso2025AGORA_plus}. The selected services span diverse application domains, including open-source platforms (e.g., GitLab and GitHub), commercial APIs (e.g., Spotify, Yelp, YouTube), and domain-specific services (e.g., Canada Holidays and OMDb).
The dataset contains a mixture of read-only (\texttt{GET}) and state-changing (\texttt{POST}) operations, to evaluate our tool's ability to infer inter-parameter constraints and workflow-dependent behavioral rules.

\subsubsection*{Setup and Metrics}
% {\color{custom-red}Execution budget (i.e., time, number of request) of each tool. Dualmine is configured to support many different LLMs and providers.}
For dynamic mining, requests are~generated until at least 50 valid (non-empty) responses are collected for each endpoint. During refinement, each non-equivalent constraint is validated with up to 30 generated counterexamples. We support multiple LLMs through a unified interface.
 
Using official API documentation and releases, we~manually evaluated and reported {\bf \em True Positives} (TP), {\bf \em False Positives} (FP), and {\bf Precision} (P). A TP is a constraint that~correctly characterizes API behavior. {\tool} could further refine TP constraints to narrow their domains (tighter ranges or additional predicates), which we refer to as {\bf \em Refined} constraints. 

%{\color{custom-blue}{
%In addition to effectiveness, we evaluate the efficiency of LLM token usage using  \emph{Constraint Efficiency}, defined as $\textit{Constraint Efficiency} = \frac{\textit{\# TP}}{\textit{Token Usage}}$. A higher value indicates better cost-effectiveness, demonstrating a method's ability to infer more correct constraints with fewer tokens.
%}}
%, resulting in more precise constraint specifications.
% We measure True Positives (TP), False Positives (FP), and Precision. 
% {\color{custom-red}{
% them gia thich
% manually
% }}
\subsection{Effectiveness (RQ1)}

\begin{table}[t]
\centering
\caption{Constraint Mining Results (RQ1)}
\label{tab:rq1-mining}
% \small
\footnotesize
\tabcolsep 1.5pt
\vspace{-6pt}
\begin{tabular}{|l|c|rrr|rrr|rrr|rrr|}
\hline
\multirow{2}{*}{\textbf{API}} & \multirow{2}{*}{\textbf{\#Op}}
&
\multicolumn{3}{c|}{\textbf{AGORA+}}
&
\multicolumn{3}{c|}{\textbf{SATORI}}
&
\multicolumn{3}{c|}{\textbf{RBCTest}}
&
\multicolumn{3}{c|}{\textbf{{\tool}}}
\\ \cline{3-14}

& &
TP & FP & P(\%)
&
TP & FP & P(\%)
&
TP & FP & P(\%)
&
TP & FP & P(\%)
\\

\hline
Canada      & 4  & 53   & 2   & 96  & 35  & 0   & 100 & 45  & 1   & 98  & 63    & 1  & 98  \\
GitLab      & 26 & 460  & 517 & 47  & 551 & 72  & 88  & 293 & 77  & 79  & 1,024 & 54 & 95  \\
GitHub      & 2  & 319  & 23  & 93  & 153 & 20  & 88  & 149 & 2   & 98  & 421   & 8  & 98  \\
OMDb        & 2  & 15   & 2   & 91  & 10  & 12  & 45  & 6   & 0   & 100 & 23    & 0  & 100 \\
Spotify     & 3  & 95   & 12  & 89   & 44  & 12  & 79  & 32  & 27  & 55  & 102   & 2  & 98  \\
Yelp        & 1  & 12   & 14  & 47  & 13  & 6   & 68  & 0   & 0   & -   & 20    & 0  & 100 \\
YouTube     & 1  & 64   & 28  & 70  & 42  & 80  & 34  & 88  & 9   & 91  & 173   & 7  & 95  \\
\hline
\textbf{Total}
            & 39 & 1019 & 597 & 63  & 848 & 202 & 81  & 613 & 116 & 84  & 1827  & 72 & 96  \\
\hline
\end{tabular}
\end{table}

%of {\tool} and three representative baselines, AGORA+, SATORI, and RBCTest, 
%We evaluate each approach using the number of correctly discovered constraints (TP), incorrectly discovered constraints (FP), and precision. 

\subsubsection{\bf \em Effectiveness Comparison}
%{\color{custom-blue}{
Table~\ref{tab:rq1-mining} presents the constraint mining results across seven REST APIs, with TP, FP, and P values averaged over five runs. Overall, {\tool} achieves a precision of \textbf{96\%} and consistently~delivers better performance than the baselines across all benchmarks. On average, it improves precision by {\textbf{33}, \textbf{15},} and \textbf{12} percentage points compared to AGORA+, SATORI, and RBCTest, respectively. These results suggest that combining specification-derived constraints with runtime-observed invariants yields more accurate constraints than relying on either source independently.
It identifies 1,827 valid constraints, outperforming AGORA+ (1019), SATORI (848) and RBCTest (613). 
%The higher yield in GitLab and GitHub stems from dense cross-field relationships, implicit semantic constraints overlooked by static mining, and additional constraints captured during runtime execution.

%This demonstrates that our hybrid approach enhances both precision and the scope of behavioral constraint discovery. The higher yield in GitLab and GitHub stems from dense cross-field relationships, implicit semantic constraints overlooked by RBCTest, and additional constraints captured during runtime execution.

%is due to the higher numbers of endpoints in those APIs and also 

%}

\paragraph{{\bf \em Qualitative analysis} on cases where {\tool}'s are better than the baselines}

The results highlight the complementary strengths of static and dynamic mining techniques. AGORA+ effectively captures constraints observed in execution traces but may miss semantic relationships that rarely appear in runtime data. In contrast, SATORI and RBCTest can infer such relationships from API specifications, but may introduce false positives when specifications are incomplete or outdated. We also make three main observations:

{\bf \em i) By integrating evidence from both sources,~our framework derives the more complete constraint}.~For~example, in the GitHub API, static mining infers~the constraint \code{gte(updated\_at}, \code{created\_at)}, while runtime mining reveals format constraints, e.g., \code{and(is\-Date\-Time(updated\_at)},\code{is\-Date\-Time(created\_at))}. The final result from {\tool} is \code{and(isDateTime(updated\_at), isDateTime(created\_at), gte(updated\_at,created\_at))}, capturing both datatype requirements and relationships. 

{\bf \em ii) Runtime analysis can uncover undocumented behavioral constraints}. For example, execution traces may reveal \code{is\-Sub\-string(name, full\_name)}, indicating that a repository's short name consistently appears within its full repository name, e.g., a namespace-qualified repository identifier. This relation may not be explicitly documented in the specification, but it reflects a stable runtime behavior useful for validation. 

{\bf \em iii) Specification mining may recover better than dynamic mining}, e.g., inferring format constraints such as 
\code{and(isURL(license.html\_url), isRegex(license.html\_\\url, '\^{}https://.*\$'))}. In contrast, runtime logs may contain few non-null \code{license.html\_url} values or insufficient URL diversity, making the regex constraint difficult to validate from executions alone. By integrating and cross-validating constraints from both sources, {\tool} preserves complementary constraints while marking weakly supported constraints for further validation or refinement, resulting in a richer and more reliable constraint set than approaches based solely on specifications or execution traces.

% \verb|and(isURL(license.html_url)|, \verb|isRegex(license.html_url, '^https://.*$'))|.

\paragraph{\bf \em Types of Constraints}

%{\color{custom-red}

% \begin{table}[t]%[!ht]
% \centering
% \tabcolsep 1.5pt
% \caption{Types of Constraints detected by {\tool} (RQ1)}
% \vspace{-3pt}
% \small
% \label{tab:mining-constraint-by-category}
% \footnotesize
% \begin{tabular}{|c|l|r!{\vrule width 1.2pt}c|l|r|}
% \hline
% \textbf{No.} & \textbf{Category/Test Oracles} & \textbf{Count} &
% \textbf{No.} & \textbf{Category/Test Oracles} & \textbf{Count} \\ \hline
% 1  & isUrl                    & 1,380 & 12 & isDate                    & 107 \\ \hline
% 2  & Template Literals        & 1,028 & 13 & Array\_Specific\_Sizes     & 53  \\ \hline
% 3  & Composite                & 865   & 14 & Nullability               & 35  \\ \hline
% 4  & Value-In-Range           & 767   & 15 & isEmail                   & 19  \\ \hline
% 5  & Value-In-Set             & 569   & 16 & isNumber                  & 11  \\ \hline
% 6  & Exists                   & 537   & 17 & ArrayElementStringLength  & 4   \\ \hline
% 7  & I/O                      & 515   & 18 & DefaultValue              & 3   \\ \hline
% 8  & N-ary atomic             & 428   & 19 & isTime                    & 3   \\ \hline
% 9  & String\_Specific\_Length & 362   & 20 & ValueNotInSet             & 2   \\ \hline
% 10 & isBoolean                & 337   & 21 & ArrayTypeOracle\_isString & 1   \\ \hline
% 11 & isDateTime               & 187   &    &                           &     \\ \hline
% \end{tabular}
% \end{table}

\begin{table}[t]%[!ht]
\centering
\tabcolsep 1.5pt
\caption{Types of Constraints detected by {\tool} (RQ1)}
\vspace{-3pt}
\small
\label{tab:mining-constraint-by-category}
\footnotesize
\begin{tabular}{|c|l|r!{\vrule width 1.2pt}c|l|r|}
\hline
\textbf{No.} & \textbf{Category/Test Oracles} & \textbf{Count} &
\textbf{No.} & \textbf{Category/Test Oracles} & \textbf{Count} \\ \hline
1  & Value-In-Range            & 406 & 10 & isBoolean                & 108 \\ \hline
2  & isUrl                     & 292 & 11 & isDateTime               & 81  \\ \hline
3  & Value-In-Set              & 266 & 12 & isDate                   & 42  \\ \hline
4  & Exists                    & 230 & 13 & Array\_Specific\_Sizes   & 22  \\ \hline
5  & Composite                 & 225 & 14 & isEmail                  & 7   \\ \hline
6  & Template Literals         & 191 & 15 & ArrayElementStringLength & 4   \\ \hline
7  & I/O                       & 181 & 16 & DefaultValue             & 3   \\ \hline
8  & N-ary atomic              & 121 & 17 & isTime                   & 2   \\ \hline
9  & String\_Specific\_Length  & 114 & 18 & ValueNotInSet            & 1   \\ \hline
\end{tabular}
\end{table}
%{\color{custom-blue}{
We classified 1,769 unique constraints mined by {\tool} from the first run into 18 test oracle categories, as shown in Table~\ref{tab:mining-constraint-by-category}. Each constraint may belong to one or more categories. The most frequent category is \code{Value-In-Range}, with 406 constraints (23\%), indicating that bounded numeric or ordinal values are commonly specified in API schemas and responses. This is followed by \code{isUrl}, containing 292 constraints (16.5\%), reflecting the prevalence of URL-related attributes, and by \code{Value-In-Set}, with 266 constraints (15\%), capturing enumerated values such as status codes, country abbreviations, and predefined identifiers. The \code{Exists} category accounts for 230 constraints (13\%), highlighting the importance of presence and nullability checks, while \code{Composite} contains 225 constraints (12.7\%), demonstrating that {\tool} can infer complex logical conditions composed of multiple atomic constraints. Template (Regex) Literals contribute 191 constraints (10.8\%), showing that regular-expression-based patterns are also frequently discovered. Moreover, the \code{I/O} category includes 181 constraints (10.2\%), indicating that {\tool} effectively captures logical relationships between input parameters and output fields. The remaining categories, including n-ary atomic constraints, string length restrictions, type-specific predicates (e.g., \code{isBoolean}, \code{isDateTime}, and \code{isDate}), array constraints, and default values, occur less frequently but showing that {\tool} covers a broad spectrum of test oracles typically found in real-world REST APIs.

\begin{table}[t]
\centering
\caption{Relationship Between Static and Dynamic Constraints (RQ1)}
\label{tab:rq1-complementarity}
\small
\tabcolsep 3pt
\vspace{-6pt}
\begin{tabular}{|l|r|rr|rrrr|}
\hline

\multirow{2}{*}{\textbf{Dataset}}
&
\multirow{2}{*}{\textbf{Eq.}}
&
\multicolumn{2}{c|}{\textbf{Unique}}
&
\multicolumn{4}{c|}{\textbf{Non Eq.}}
\\ \cline{3-8}

&
&
\textbf{Static}
&
\textbf{Dynamic}
&
\textbf{Subset}
&
\textbf{Inter.}
&
\textbf{Indep.}
&
\textbf{Union}
\\

\hline

Canada      & 3    & 34.6   & 9.2   & 8.6   & 8.4   & 0 & 0   \\
GitLab      & 58.2 & 611.6  & 164.6 & 42    & 183.6 & 3 & 15  \\
GitHub      & 46.8 & 176.4  & 76.4  & 23.2  & 105.4 & 0 & 0.8 \\
OMDb        & 1.2  & 19.8   & 1.2   & 0.6   & 0.6   & 0 & 0   \\
Spotify     & 8.8  & 50.6   & 11.2  & 17    & 16.2  & 0 & 0.2 \\
Yelp        & 2.8  & 11.2   & 1.8   & 1.2   & 3.4   & 0 & 0   \\
YouTube     & 5    & 134.4  & 16    & 11    & 14    & 0 & 0   \\

\hline

\textbf{Total}
& 125.8
& 1038.6
& 280.4
& 103.6
& 331.6
& 3
& 16
\\

\hline
\end{tabular}
\end{table}

\subsubsection{\bf \em Sources of Constraints}

% Table~\ref{tab:rq1-complementarity} analyzes the relationship between constraints discovered from static and dynamic mining. We categorize the discovered constraints into four groups: equivalent constraints identified by both approaches (Eq.), constraints unique to either static or dynamic mining (Unique), and non-equivalent relationships including subset, intersection, disjoint, and union relationships.
Table~\ref{tab:rq1-complementarity} analyzes the relationship between the constraints discovered from {\bf static} and {\bf dynamic} mining after validation. From the average of 1,827 TP and 72 FP constraints after five runs, we categorized the constraints into equivalent ones identified by both approaches (Eq.), constraints unique to either static or dynamic mining (Unique), and non-equivalent constraints including subset, intersection, independence, and union relationships.

\paragraph{\bf \em Unique Constraints/Invariants}

% The results show that the overlap between the two sources is relatively limited. Although a number of equivalent constraints are consistently discovered by both approaches, a substantial portion of the constraints are unique to either static or dynamic mining. This finding indicates that the two mining strategies capture different aspects of API behavior. Static mining is effective at extracting semantic constraints explicitly or implicitly described in API specifications, whereas dynamic mining excels at uncovering behavioral relationships manifested in runtime executions.

% Example:...
The results demonstrate strong complementarity between static and dynamic mining. Across all datasets, only 125.8 constraints are equivalent, while  1,038.6  and 280.4 constraints are uniquely discovered by static and dynamic mining, respectively. Unique static constraints come from schemas, descriptions, formats, and documented rules. Unique dynamic constraints come from observed runtime relations, concrete field correlations, and execution-specific behaviors. The imbalance does not mean static is better; it may reflect that LLM/static mining produces many more candidate constraints. This also indicates that the two approaches capture different aspects of API~behavior.

Static mining primarily derives semantic constraints, type definitions, and parameter restrictions from specifications, whereas dynamic mining uncovers behavioral invariants from execution traces. For example, static analysis may infer constraints such as \code{gte(updated\_at, created\_at)}, while runtime analysis reveals undocumented properties that are absent from the specification, including \code{isSubstring(name, full\_name)}, \code{isNumeric(totalResults)} despite \code{totalResults} being represented as a string, \code{isURL(return.pulls\_url)}, and \code{eq(sizeOf(return.items[].id), 11)} in Youtube API. The latter captures a domain-specific invariant of the YouTube platform, where system-generated video identifiers consistently contain exactly 11 characters.

\paragraph{\bf \em Subset and Intersection}

% Furthermore, the presence of subset and intersection relationships suggests that many constraints inferred from one source partially overlap with those discovered from the other source rather than being exact matches. Disjoint relationships reveal cases where the two approaches capture completely different behavioral properties. 

% Example:...
A large number of constraints exhibit 103.6 subset and 331.6 intersection relationships, indicating that static and dynamic mining often discover partially overlapping rather than identical constraints.

Subset relationships occur when one source derives a stricter version of a constraint discovered by the other. For example, dynamic mining infers \code{isURL(return.owner.url)}, while static mining derives 
\code{\texttt{and(isURL(return.owner.url),}
\code{isRegex(return.owner.url, \string^https://))}}, adding~an~`https' requirement. 
In contrast, intersection relationships arise when static and dynamic mining capture different but compatible properties. For example, static mining infers~\code{is\-URL\- (return\-.url)}, whereas dynamic mining discovers~\code{is\-Substring(return.url, return.releases\_url)}. Neither constraint subsumes the other, yet both describe valid aspects of the field behavior. The large number of intersection cases highlights the complementary nature of two approaches. 

\paragraph{\bf \em Independence set}
Only 3 independent relationships are observed, all of which are from GitLab 
APIs. This~small number indicates that direct contradictions between static and dynamic constraints are relatively rare.
Independent relationships arise when static and dynamic mining infer completely different constraints. For example, static mining derives URL-format constraints on \code{return.web\_url}, while dynamic mining discovers \code{isSubstring(input.branch, return.web\_url)}. Since the {\bf \em two complementary constraints} describe unrelated properties, they form a independent relationship, illustrating the distinct perspectives provided by specification and runtime~analysis.

\paragraph{\bf \em Union set}
We identify 16 union relationships, also primarily in GitLab. These occur when static and dynamic mining infer complementary constraints that can be combined into a richer representation. 
For example, for the GitLab \code{GET /projects} endpoint, static mining infers the conditional constraint \code{implies(toBool(default(input.with\_issues\_e\-nabled,`false')), eq(issues\_enabled,true))},~while~dy\-namic mining infers the value-domain constraint \code{in(issues\_enabled,[0,1])}. Since the two constraints~capture complementary aspects of the API behavior, their union yields the stronger specification.

% {\color{custom-red}A static constraint may infer \code{isURL(return.url)}, while dynamic mining infers \code{isRegex(return.url, '\^{}https://api.github.com/.*')}.
% Their union yields a stronger constraint combining general URL validity
% with the observed API-domain pattern. OR OTHER EXAMPLE?} 

%{\color{custom-red} Any other example???}
%For example, static mining may derive \code{gte(updated\_at, created\_at)}, while dynamic mining discovers \code{isDateTime(updated\_at)} and \code{isDateTime(created\_at)}. Their union yields \code{and(isDateTime(updated\_at), isDateTime(created\_at), gte(updated\_at, created\_at))}, capturing both datatype and semantic properties.

%{\color{custom-blue}
%\paragraph{Error Analysis of {\tool}}

\subsubsection{\bf \em Error Analysis of {\tool}}
We perform a qualitative analysis of the false-positive constraints generated by {\tool} to characterize its limitations. Most false positives occur in the \textit{Unique} category, i.e., constraints mined from only one source without cross-source support. This suggests that dual validation is most effective when static and dynamic evidence can corroborate or challenge each other, but false positives may still remain when a constraint is supported by only one source and no decisive counter-evidence is available.

The false positives produced by {\tool} mainly fall into two categories. {\bf First}, \textbf{spurious cross-field correlations} arise when fields representing unrelated domain concepts accidentally satisfy a mathematical or logical relation in the observed traces. For example, in the GitLab Issue API, {\tool} inferred \code{gte(assignee.id, size(labels[]))}, suggesting that the assignee identifier is always greater than or equal to the number of labels. This relation is not semantically meaningful: database-generated identifiers tend to be large, while label arrays usually have low cardinality, causing the inequality to hold accidentally in the collected traces. Similarly, in the GitHub API, {\tool} inferred \code{lte(allow\_auto\_merge, allow\_forking)} between two independent repository settings. Under a boolean encoding where \code{false < true}, this constraint implies that auto-merge can be enabled only when forking is enabled. The relation held in the observed traces but is not a general API invariant.

{\bf Second}, \textbf{trace overfitting} occurs when the execution logs cover only a narrow subset of valid database states, user permissions, or environment configurations. For example, in the GitHub API, the collected traces lacked diverse organization and collaborator configurations, leading {\tool} to infer that organization members never have push permission, i.e., \code{permissions.push}==\code{false}. Although this property held for the specific organization scope during testing, it is violated in repositories with more permissive settings. Such false positives reflect sampling bias rather than semantic API rules.

These errors indicate that the limitations of {\tool} are mainly caused by insufficient evidence for constraints that are unique to one source. They suggest directions for improvement. First, the validator can incorporate stronger semantic checks to reject relations between fields with incompatible domain roles, such as identifiers and collection sizes. Second, the counterexample generator could target underrepresented configurations, such as alternative permission levels, repository settings, and organization states, to reduce trace~overfitting.

\subsection{Dual-Validation Capability (RQ2)}

To assess the contribution of our dual-validation, we compare the complete version of {\tool} (\textit{Full \tool}) with a variant in which validation is disabled (\textit{w/o DV}). 
%Without dual validation, all constraints generated during the hybrid mining stage are retained directly, whereas the full version performs additional semantic and execution-based verification before accepting a constraint. 
%Table~\ref{tab:ablation-study} reports the resulting numbers of true positives (TP), false positives (FP), and precision.

As seen in Table~\ref{tab:ablation-study}, dual validation substantially improves constraint quality. In one of five runs across all datasets, the full version achieves a precision of 96\%, compared with 83\% without validation. Although disabling validation increases the number of true positives (2,264 vs. 1,789), it also introduces nearly six times more false positives (479 vs. 81). %Similar trends are observed across most APIs, e.g., GitLab, GitHub, Spotify, and YouTube.

The differences in TP and FP show the tradeoff introduced by validation. Overall, dual validation removes 398 false positives. 
Thus, dual validation trades a 21.9\% reduction in valid-constraint yield for an 82.9\% reduction in false positives, increasing precision from 83\% to 96\%.
%corresponding to an 82.9\% FP reduction, while also removing 475 true positives, corresponding to a 21.9\% reduction in valid constraints. 
This indicates that validation is conservative: it substantially improves precision, but may discard or leave unresolved some valid constraints when complementary evidence is insufficient.

%Overall, dual validation reduces false positives by 82.9\% (479 to 82) and improves precision from 83\% to 96\%. This improvement comes with a reduction in true positives from 2,264 to 1,769, indicating that validation is intentionally conservative when evidence is insufficient.

This result supports the design of dual validation. By checking mined constraints against complementary evidence from specifications and executions, {\tool} removes~document\-ation-induced and trace-induced false positives. 
Nevertheless, the full version retains 78.1\% of the valid constraints while eliminating most false positives. This tradeoff 
%improves the reliability of the final constraints and 
also motivates the~counter\-example-guided refinement, which aims to specialize valid constraints that remain unresolved after~validation.

The benefit of dual validation varies across APIs. The largest gain occurs on GitLab, where false positives drop from 358 to 52 and precision improves from 79\% to 95\%. This suggests that GitLab produces many plausible but unsupported constraints, likely because its richer repository metadata creates more opportunities for both documentation-induced and trace-induced false positives. Dual validation also improves precision on GitHub, Spotify, Yelp, and YouTube, showing that the effect is not limited to one API.

\begin{table}[t]
\centering
\caption{Ablation Study of Dual Validation (DV) (RQ2) and Counterexample (CE)-Guided Refinement. (RQ3)}
%\caption{Ablation Study (RQ2). DV: dual-validation; CE: counterexamples; Refined: properties with better constraints ranges in Full vs. w/o CE.}
\label{tab:ablation-study}
\footnotesize
\tabcolsep 2.5pt
\vspace{-5pt}
\begin{tabular}{|l|rrr|rrr|rrr|c|}
\hline
\multirow{2}{*}{Datasets}
&
\multicolumn{3}{c|}{Full \tool} &
\multicolumn{3}{c|}{w/o DV} &
\multicolumn{3}{c|}{w/o CE} &
\multirow{2}{*}{Refined} 
\\ \cline{2-10}
&
TP & FP & P (\%) &
TP & FP & P (\%) &
TP & FP & P (\%) & 
\\
\hline
Canada      & 64    & 0  & 100 & 62    & 4   & 94  & 63    & 1  & 98  & 0  \\
GitLab      & 1,009 & 50 & 95  & 1,357 & 358 & 79  & 1,009 & 50 & 95  & 30 \\
GitHub      & 399   & 17 & 96  & 473   & 75  & 86  & 398   & 18 & 96  & 35 \\
OMDb        & 22    & 0  & 100 & 25    & 0   & 100 & 22    & 0  & 100 & 0  \\
Spotify     & 102   & 1  & 99  & 120   & 11  & 92  & 102   & 1  & 99  & 13 \\
Yelp        & 20    & 0  & 100 & 32    & 2   & 94  & 20    & 0  & 100 & 1  \\
YouTube     & 173   & 13 & 93  & 195   & 29  & 87  & 174   & 12 & 94  & 9  \\
\hline
Total       & 1,789 & 81 & {\bf 96}  & 2,264 & 479 & {\bf 83}  & 1,788 & 82 & 96  & {\bf 88} \\
\hline
\end{tabular}

%\vspace{6pt}
%Em thêm câu này làm reader confuse là tại sao như vậy.
%\footnotesize{\textit{Note: Only the first run of each dataset is used for comparison}.}

\end{table}

Most rejected constraints stem from limitations of the individual mining sources. Specification-derived constraints may originate from incomplete or outdated documentation that does not accurately reflect the implementation, while runtime-derived constraints may result from limited execution diversity, causing coincidental patterns in traces to be misidentified as genuine invariants. For example, runtime mining may infer \code{contains(["Apache License 2.0", "MIT License"], return[].license.name)} for \code{GitHub\_GetOrganizationRepositories} because all observed repositories happen to use one of these licenses. However, GitHub repositories can be distributed under many other licenses, making the inferred constraint a sampling artifact.

\subsection{Counterexample (CE)-Guided Refinement (RQ3)}
%{\color{custom-blue}{
To evaluate the contribution of counterexample-guided refinement, we compare the full version of {\tool} with a variant that disables CE while keeping hybrid mining and dual validation unchanged. Table~\ref{tab:ablation-study} shows that CE has little effect on coarse TP/FP-based correctness metrics. The full version and the \textit{w/o CE} variant both achieve 96\% precision, differing by only one true positive and one false positive across all datasets. This result indicates that the dual-validation stage already removes more incorrect constraints before refinement.

%To evaluate the effectiveness of the proposed counterexample-guided refinement mechanism, Table~\ref{tab:ablation-study} presents an ablation study. Removing CE has little impact on detection performance: both the Full Dual Mine and the \textit{w/o CE} variants achieve 96\% precision, differing by only one TP/FP across all datasets. This indicates that the dual-validation stage already eliminates most incorrect constraints.

The main contribution of CE is therefore not to increase the number of accepted constraints, but {\bf \em to improve the specificity and expressiveness} of the accepted constraints. By generating targeted counterexamples for non-equivalent or uncertain constraints, CE can strengthen overly general constraints, add missing predicates, and specialize constraints to more precise behavioral conditions. Overall, as shown in the Refined column of Table~\ref{tab:ablation-study}, CE improves the specificity of 88 constraints, with the largest numbers on GitHub (35), GitLab (30), Spotify (13), and YouTube (9). These results show that TP/FP alone underestimates the value of CE because two constraints may both be counted as true positives even when one is more precise and useful as a test oracle.

%The primary contribution of CE is improving constraint quality rather than increasing the number of accepted constraints. By generating targeted counterexamples and re-validating non-equivalent constraints, CE refines overly permissive constraints by tightening boundaries and adding missing predicates. Overall, CE refines 88 properties, with the largest improvements on GitHub (35), GitLab (30), Spotify (13), and YouTube (9).

For the GitLab endpoint \code{GET} \code{/projects/id/repository/} \code{branches}, the \textit{w/o CE} variant infers only: \code{isSubstring} \code{(commit.short\_id, commit.id)}. After counterexample-guided refinement, the full version strengthens this constraint by adding the missing SHA-format predicates for both fields:
\begin{quote}
\footnotesize
\begin{verbatim}
and(
isRegex(commit.id, '^[a-f0-9]{40}$'),
isRegex(commit.short_id, '^[a-f0-9]{7,8}$'),
isSubstring(commit.short_id, commit.id)
)
\end{verbatim}
\end{quote}
Thus, CE converts a valid but incomplete relationship into a more informative oracle that captures both the containment relation and the expected commit-hash formats. 
%As seen in~Fig. ~\ref{fig:architecture}, this iterative refinement improves constraint specificity while preserving high precision achieved by dual validation.

The results also reveal a limitation. Because CE refines constraints using additional generated executions, it may occasionally introduce or retain a constraint that is later judged false, as reflected by the small FP difference on YouTube. That is, {\em the refined constraints might not always be correct}. However, this effect is minor in aggregate: the full version preserves the same overall precision as \textit{w/o CE} while producing substantially more precise constraints for 88 properties.~Overall, {\bf \em CE complements dual validation by improving constraint quality}.

\subsection{Cost-Effectiveness Analysis (RQ4)}

\begin{table}[t]
\centering
\caption{Token Consumption and Constraint Efficiency (RQ4)}
\label{tab:rq4-token-efficiency}
\footnotesize
\tabcolsep 2.5pt
\vspace{-6pt}
\begin{tabular}{|l|rrr|rrr|rrrr|}
\hline

\multirow{2}{*}{\textbf{API}}
&
\multicolumn{3}{c|}{\textbf{SATORI}}
&
\multicolumn{3}{c|}{\textbf{RBCTest}}
&
\multicolumn{4}{c|}{\textbf{\tool}}
\\ \cline{2-11}

&
\textbf{In}
&
\textbf{Out}
&
\textbf{Eff.}
&
\textbf{In}
&
\textbf{Out}
&
\textbf{Eff.}
&
\textbf{In}
&
\textbf{Out}
&
\textbf{Cache}
&
\textbf{Eff.}
\\

\hline

Canada      & 27.8 & 3.7 & 1.1 &  20.0 &   3.1 & 1.9 & 497.2 &  23.9 &   4.0 & 0.1 \\
GitLab      & 458 & 66.4 & 1 & 630.3 &  88.5 & 0.4 & 179.5 & 340.7 &  26.1 & 1.9 \\
GitHub      & 298.3 & 44.9 & 0.4 & 170.8 &  17.2 & 0.8 & 120.1 & 117.4 &  50.1 & 1.5 \\
OMDb        & 16.7 & 2.6 & 0.5 &  29.0 &   3.3 & 0.2 &   4.3 &   1.4 &   1.5 & 3.3 \\
Spotify     & 42.7 & 5.9 & 0.9 &  74.0 &   8.1 & 0.4 &  21.7 &  16.0 &   6.2 & 2.3 \\
Yelp        & 21 & 2.7 & 0.5&   1.4 &   0.2 & 0.0 &  12.1 &  12.2 &  12.3 & 0.7 \\
YouTube     & 102.4 & 16.5 & 0.4 & 171.2 &  21.3 & 0.5 & 139.8 & 150.4 &  22.9 & 0.5 \\

\hline

\textbf{Total}
& 966.8
& 144.7
& 0.8
& 1096.5 & 141.6 & 0.5 & 974.7 & 662.0 & 123.1 & 1.1
\\

\hline
\end{tabular}
% Don't remove the space between the below text and figure, make it look ugly.
%\vspace{-16pt}
\end{table}

%{\color{custom-red}{Additional with SATORI Comment}}

To evaluate the efficiency of LLM token usage, we define $\textit{Constraint Efficiency} = \frac{\textit{\# TP}}{\textit{(Token Usage)/1k}}$. A higher value indicates better cost-effectiveness, demonstrating a method's ability to infer more correct constraints with fewer tokens.

Table~\ref{tab:rq4-token-efficiency} reports token consumption and constraint efficiency for the approaches, measured as the number of validated constraints discovered per 1k consumed tokens. We exclude AGORA+ because it does not use LLMs. Overall, {\tool} achieves higher efficiency than RBCTest (1.1 versus 0.5 constraints per 1k tokens, a 2.2$\times$ improvement) and SATORI (1.1 versus 0.8 constraints per 1K tokens). This indicates~that {\tool} discovers more validated constraints for a comparable token budget.
Although it produces more output tokens (662.0K vs. 141.6K and 144.7), it uses fewer input tokens (974.7K vs. 1096.5K and 966.8). The additional output cost mainly comes from validation and counterexample-guided refinement, which iteratively analyze and revise candidate constraints. Despite this overhead, these stages improve constraint quality and yield higher efficiency. The gains are especially large on GitLab (1.9 vs. 0.4 and 1), OMDb (3.3 vs. 0.2 and 0.5), and Spotify (2.3 vs. 0.4 and 0.9). While {\tool} is less efficient on a few APIs such as Canada, the overall results show that investing tokens in validation produces substantially more correct constraints.

In terms of monetary cost, DualMine requires about \$7.85, compared with \$2.79 for RBCTest and \$2.66 for SATORI. Thus, DualMine is 2.8× more expensive than them. However, the higher cost is primarily driven by the validation and refinement phases, which consume additional output tokens to verify candidate constraints and generate counterexamples. This extra investment leads to substantially higher constraint correctness and efficiency across most APIs, indicating that the increased token cost is a trade-off that improves the reliability and practical usefulness of the mined constraints~\footnote{GPT-5 pricing: \$1.25/1M input tokens (\$0.125 cached) and \$10/1M output tokens.}

\subsection{Fault Detection (RQ5)}
%\vspace{-3pt}
%\subsubsection{Procedure} 

We study whether the constraints discovered by {\tool} can serve as effective test oracles for detecting real API faults. We define {\em a fault as a contract violation} in which an API response contradicts a validated behavioral constraint supported by the specification, official documentation, repeated executions, or source-level evidence when available. To avoid counting duplicate reports, we group repeated violations that share the same endpoint, response field, violated constraint, and root cause into a single fault. 
We distinguish two sources of fault discovery. In \textit{validation-time fault discovery}, a specification-supported constraint is violated by an observed response during dual validation. Such a violation is not immediately treated as a false constraint. Instead, we manually inspect the violated constraint, the observed response, the API documentation, and repeated executions to determine whether the violation reflects an invalid constraint or a real contract violation. In \textit{oracle-time fault discovery}, a constraint that has already passed validation is used as a test oracle on newly executed requests. Any response that violates such a validated constraint is reported as a candidate fault and manually confirmed using the same evidence. We report de-duplicated faults across both sources.

%Fault detection follows a two-stage process integrated with the dual-validation framework. First, during validation, a specification-derived constraint may be violated by observed executions. Such a violation is not immediately treated as a false constraint. Instead, we manually inspect the violated constraint. Only violations supported by external evidence are counted as faults. Second, after constraint aggregation and refinement, the validated constraints form the final test-oracle set. We re-execute API requests and report any violation of these validated constraints as a potential fault, which is then manually confirmed in the same procedure.

%Fault detection follows a two-stage process integrated with the dual-validation framework. During validation, any constraint that fails validation is manually analyzed to determine if the failure is due to an implementation defect rather than an invalid constraint. Confirmed cases are reported as faults. After constraint aggregation, the validated constraints form the final set of test oracles. API requests are then re-executed, and any violation of these validated constraints is reported as a fault.

%\subsubsection{Empirical Results}

% Figure~\ref{fig:bug} shows the distribution of the 48 faults detected by \tool in eight validation-constraint types. \code{DefaultValue} is the most frequent (33.3\%), followed by \code{Template Literals} (26.7\%), indicating that incorrect default values and source-level programming errors are the dominant fault patterns.
%{\color{custom-blue}{

\begin{figure}[t]
  \centering
  \includegraphics[width=3.3in]{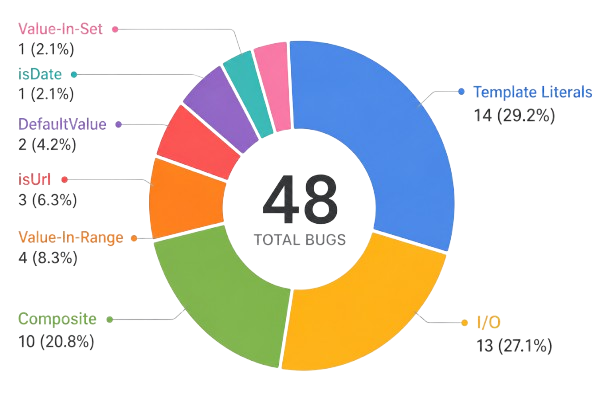}
  \vspace{-15pt}
  \caption{{\tool}: Bug Category Distribution}
  \label{fig:bug}
\end{figure}

Fig.~\ref{fig:bug} shows the distribution of the 48 faults in 8 constraint types detected~by~{\tool}. \code{Template}~\code{Literals} is the most frequent (29.2\%), followed by \code{I/O} (27.1\%)~and \code{Composite} (20.8\%), indicating that regular-expression errors, implementation-level I/O errors, and composite validation logic account for the majority of detected faults.
The \code{DefaultValue} category captures inconsistent default-value behavior, e.g., \code{anonymous\_access\_enabled} is specified with a default value of \code{true} but the API returns \code{null}. The \code{isUrl} and \code{isNumber} categories detect contract violations, such as the OMDb API returning \code{"N/A"} for \code{Poster} and \code{Metascore} despite their \code{isUrl} and \code{isNumber} constraints. The \code{Composite} category identifies violations across multiple fields, e.g., \code{updated\_at >= created\_at}. Other categories, including \code{Value-In-Range}, \code{isUrl}, \code{DefaultValue}, \code{isDate}, and \code{Value-In-Set}, capture invalid ranges, malformed values, and external-resource failures.
%}}
% The \code{DefaultValue} category captures inconsistent default-value behavior, e.g., \code{anonymous\_access\_enabled} is specified with a default value of \code{true} but the API returns \code{null}. The \code{isUrl} and \code{isNumber} categories detect contract violations, such as the OMDb API returning \texttt{"N/A"} for \code{Poster} and \code{Metascore} despite their \code{isUrl} and \code{isNumber} constraints. The \code{Composite} category identifies violations across multiple fields, e.g., the GitLab constraint \code{last\_activity\_at >= created\_at}. Other categories, including \code{Value-In-Range}, \code{isDate}, and \code{I/O}, capture invalid ranges, malformed temporal values, and external-resource failures.

%Overall, \code{Template Literals} represents static faults detectable through source-code analysis, whereas the remaining categories mainly correspond to dynamic faults requiring validation of runtime inputs or API responses. These results show that \tool\ primarily targets runtime contract violations beyond the reach of conventional static analysis.

\subsection{Threats to Validity}
\label{sec:threats_to_validity}

%\textbf{Construct Validity.}
%We use standard metrics (TP, FP, precision, and constraint efficiency) and manually validated ground-truth constraints from the AGORA+ and RBCTest benchmarks. Although we carefully verified the labels, minor annotation inaccuracies may remain.

\textbf{Internal Validity.}
The results may be affected by LLM configurations, prompts, and implementation choices. To mitigate this threat, we used a consistent setup across all APIs and validated our implementation and baseline integrations. 

\textbf{External Validity.}
We evaluated {\tool} on diverse real-world REST APIs from the AGORA+ and RBCTest benchmarks. Although the dataset spans multiple domains, the findings may not generalize to atypical APIs, especially those with sparse documentation or limited execution traces.

\textbf{Conclusion Validity.}
The gains may depend on the selected benchmark, specifications, and test workloads. To strengthen validity, we compared \tool against multiple baselines.
%and evaluated both effectiveness and costs.

%\textbf{Construct Validity.}
%Our evaluation employs standard metrics (TP, FP, precision, and constraint efficiency), utilizing manually validated ground-truth constraints from the AGORA+ and RBCTest benchmarks. While we conducted rigorous verification, minor labeling inaccuracies inherent to manual annotation may still persist.

%\textbf{Internal Validity.}
%The results may be affected by LLM configurations, prompting strategies, and implementation choices. To mitigate this threat, we used a consistent experimental setup across all APIs and carefully validated our implementation and baseline integrations.

%\textbf{External Validity.} We evaluated our approach using a highly diverse set of real-world REST APIs from the AGORA+ and RBCTest benchmark. While this dataset spans multiple domains and ensures broad representativeness, our findings may not fully generalize to atypical APIs—particularly those with exceptionally sparse documentation or a lack of execution traces.

%\textbf{Conclusion Validity.}
%The observed improvements may depend on the selected benchmark, specifications, and test workloads. To strengthen validity, we compared {\tool} against multiple baselines and evaluated both effectiveness and cost-related metrics.

\section{Related Work}
\label{sec:related}

% Recent surveys~
Prior surveys and studies highlight the growing automation in REST API testing~\cite{kim2022automated-2,golmohammadi2022testing-2,ehsan2022restful,martin2022online-2,sharma2018automated,marculescu2022faults-2,martin2021black-2, molina2025oracle}. Existing approaches are broadly divided into response validation, oracle generation, and AI-assisted API testing.
% Existing approaches can be broadly divided into response validation, oracle generation, and AI-assisted API~testing.

\textbf{Response Validation.}
Status-code validation checks if an API returns the expected HTTP status and is widely used in Postman \cite{postman}, Katalon~\cite{katalon}, RestTestGen~\cite{viglianisi2020resttestgen-2}, and KAT~\cite{kat-2}. Schema validation checks required fields and data types but cannot verify semantic correctness, so responses may pass validation while still being logically incorrect.

\textbf{Oracle and Constraint Mining.}
Prior REST API oracle research has explored metamorphic relations~\cite{segura2018metamorphic}, mining-based ~\cite{alonso2025satori,huynh2026rbctest,alonso2023agora,Alonso2025AGORA_plus}, and learning-based~\cite{dinella2022toga, wu2024lam4inv, chakraborty2023ranking}, enabling the inference of richer correctness constraints beyond structural checks.
% Prior REST API oracle research has explored metamorphic relations~\cite{segura2018metamorphic}, static mining~\cite{huynh2026rbctest,alonso2025satori}, dynamic mining~\cite{alonso2023agora, Alonso2025AGORA_plus} and neural oracle \cite{dinella2022toga}, enabling the inference of richer correctness constraints beyond structural checks.
EmRest uses error-message evidence for REST API testing~\cite{xu2025emrest}. Static REST analysis tools (e.g., ReSpecTor and REST$\pi$) recover specifications or path-sensitive API types from implementations~\cite{huang2024respector,aldrich2025restpi}.

% \cite{godefroid2005dart, majumdar2007hybrid, csallner2008dysy}
\textbf{AI-Assisted API Testing.}
ML/LLMs have been used for API input generation, dependency reasoning, graph-based exploration, and specification enhancement~\cite{gu2026panta, godefroid2005dart,majumdar2007hybrid, csallner2008dysy, alonso2022arte,atlidakis2019restler-2,liu2022morest-2,kim2024leveraging,kim2025autoresttest,kim2025LlamaRestTest,wen2024autospec,saha2025raft}. These methods improve test generation and exploration, whereas we focus on discovering response-body constraints as test oracles. {\tool} leverages APIPilot~\cite{api_pilot} as the execution engine to automatically produce diverse execution traces and counterexamples. These runtime artifacts support dynamic constraint mining and iterative validation.

\section{Conclusion}

This paper presented {\tool}, a hybrid framework for automated REST API constraint discovery that combines specification-based mining, runtime invariant mining, dual validation, and counterexample-guided refinement. By leveraging both static and dynamic evidence, {\tool} discovers more valid constraints while achieving higher precision than existing approaches. Our results and ablation studies demonstrate the effectiveness of the proposed techniques in improving the quality of discovered API constraints.

%\subsubsection*{Data Availability Statement}
\vspace{1pt}
{\bf Data Availability Statement}. Our data and code are publicly available at our website~\cite{dual_mine_replication_2026}.

\balance

\bibliographystyle{IEEEtran}

\bibliography{references}

\end{document}